\documentclass{moriond}

\def\be{\begin{equation}}
\def\ee{\end{equation}}
\def\bea{\begin{eqnarray}}
\def\eea{\end{eqnarray}}

\newcommand{\Photo}{\includegraphics[height=35mm]{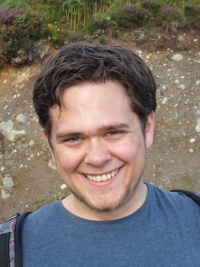}}
\newcommand{\tn}[1]{\textnormal{#1}}

\newcommand{\abs}[1]{\left| #1 \right|}

\newcommand{\GeVc}[0]{\tn{ GeV}/c}

\newcommand{\snn}{{\sqrt{s_\mathrm{NN}}}}

\newcommand{\pT}{\ensuremath{p_\mathrm{T}}}

\begin{document}
\vspace*{4cm}
\title{Charged jets in p--Pb collisions at $\snn = 5.02$ TeV measured with the ALICE detector}

\author{ R\"udiger Haake }

\address{CERN,\\
CH-1211 Geneva 23, Switzerland}

\maketitle\abstracts{Highly energetic jets are sensitive probes for the kinematics and the topology of nuclear collisions.
Jets are collimated sprays of charged and neutral particles, which are produced in the fragmentation of hard scattered partons in the early stage of the collision. The measurement of nuclear modification of charged jet spectra in p--Pb collisions provides an important way of quantifying the effects of cold nuclear matter in the initial state on jet production, fragmentation, and hadronization.
Unlike in Pb--Pb collisions, modifications of jet production due to hot nuclear matter effects are not expected to occur in p--Pb collisions. Therefore, measurements of nuclear modifications in charged jet spectra in p--Pb collisions (commonly known as $R_\mathrm{pPb}$) can be used to isolate and quantify cold nuclear matter effects.
Potential nuclear effects are expected to be more pronounced in more central p--Pb collisions due to a higher probability of an interaction between the proton and the lead-nucleus. To measure the centrality dependence of charged jet spectra, it is crucial to use a reliable definition of event centrality, which ALICE developed utilizing the Zero-Degree Calorimeter (ZDC). }

%___________________________________________________________________________________________
\section{Introduction}

Jets can conceptually be described as the final state produced in a hard scattering. Therefore, jets are an excellent tool to access a very early stage of the collision. The jet constituents represent the final state remnants of the fragmented and hadronized partons that were scattered in the reaction. While all the detected particles have been created in a non-perturbative process (i.e. by hadronization), ideally, jets represent the kinematic properties of the originating partons. Thus, jets are mainly determined by perturbative processes due to the high momentum transfer between two partons and the cross sections can be calculated with pQCD. This conceptual definition is descriptive and very simple, the technical analysis of those objects is complicated though.\\

This article presents minimum bias and centrality-dependent results for charged jets measured in proton--lead collisions at $\snn = 5.02$ TeV with the ALICE experiment. Due to lack of space, the focus will be on the presentation of the results, leaving out most technical details that are summarized very briefly in a single section.
A detailed description of the analyses presented in these proceedings can be found in the respective publication~\cite{MinBiasPaper}~\cite{CentralityPaper}.

%___________________________________________________________________________________________
\section{Experimental details}
\label{sec:Details}

% ALICE
The data were recorded with ALICE, the dedicated heavy-ion experiment at the LHC studying properties of the quark-gluon plasma and the QCD phase diagram in general. The detector is designed as a general-purpose heavy-ion detector~\cite{ALICE2008} to measure and identify hadrons, leptons, and also photons down to very low transverse momenta.

For the charged jet analysis, mainly data from the ALICE TPC~\cite{ALICE2010b} -- a time projection chamber --, and the ITS~\cite{ALICE2010} -- a six-layered silicon detector -- is used to form charged-particle tracks that serve as the basic ingredient to jet reconstruction.
For event triggering, the VZERO~\cite{ALICE2010c} scintillation counters are utilized. The centrality estimation method makes use of the Zero-Degree Calorimeter (ZDC), a quartz fibers sampling calorimeter located 116 m from the interaction point. Details on the centrality selection can be found here~\cite{CentralityMethodPaper}.\\

% Jet reconstruction

To identify jets, the anti-$k_\mathrm{T}$ algorithm~\cite{Cacciari2008} implemented in the FastJet \cite{Cacciari2006} package is used. The track cuts for those particles are chosen in order to obtain a uniform charged track distribution in the full $\eta-\phi$ plane and only tracks with $\abs{\eta} < 0.9$ and $\pT > 150 \mbox{ MeV/}c$ are accepted for the jet finding procedure.

To avoid edge effects, only jets fully contained within the acceptance are used for further analysis.\\

% Correction methods

Two corrections are applied to the raw jets after reconstruction: background correction, which includes the subtraction of the mean event background density and the consideration of the background fluctuations within the event, and the correction for detector effects.

While the background density is subtracted on a jet-by-jet basis, region-to-region fluctuations of the background are measured by probing the transverse momentum in randomly distributed cones and comparing it to the average background. While the background subtraction can be applied for each jet, the background fluctuations can only be taken into account on a probabilistic basis in an unfolding procedure.

Like the background fluctuations, detector effects -- e.g. from the limited single-particle tracking efficiency -- are considered in a Singular Value Decomposition (SVD) unfolding procedure~\cite{SVDAlgorithm}.

%___________________________________________________________________________________________
\section{Results}
\label{sec:Results}

\begin{figure}
\includegraphics[width=1.0\textwidth]{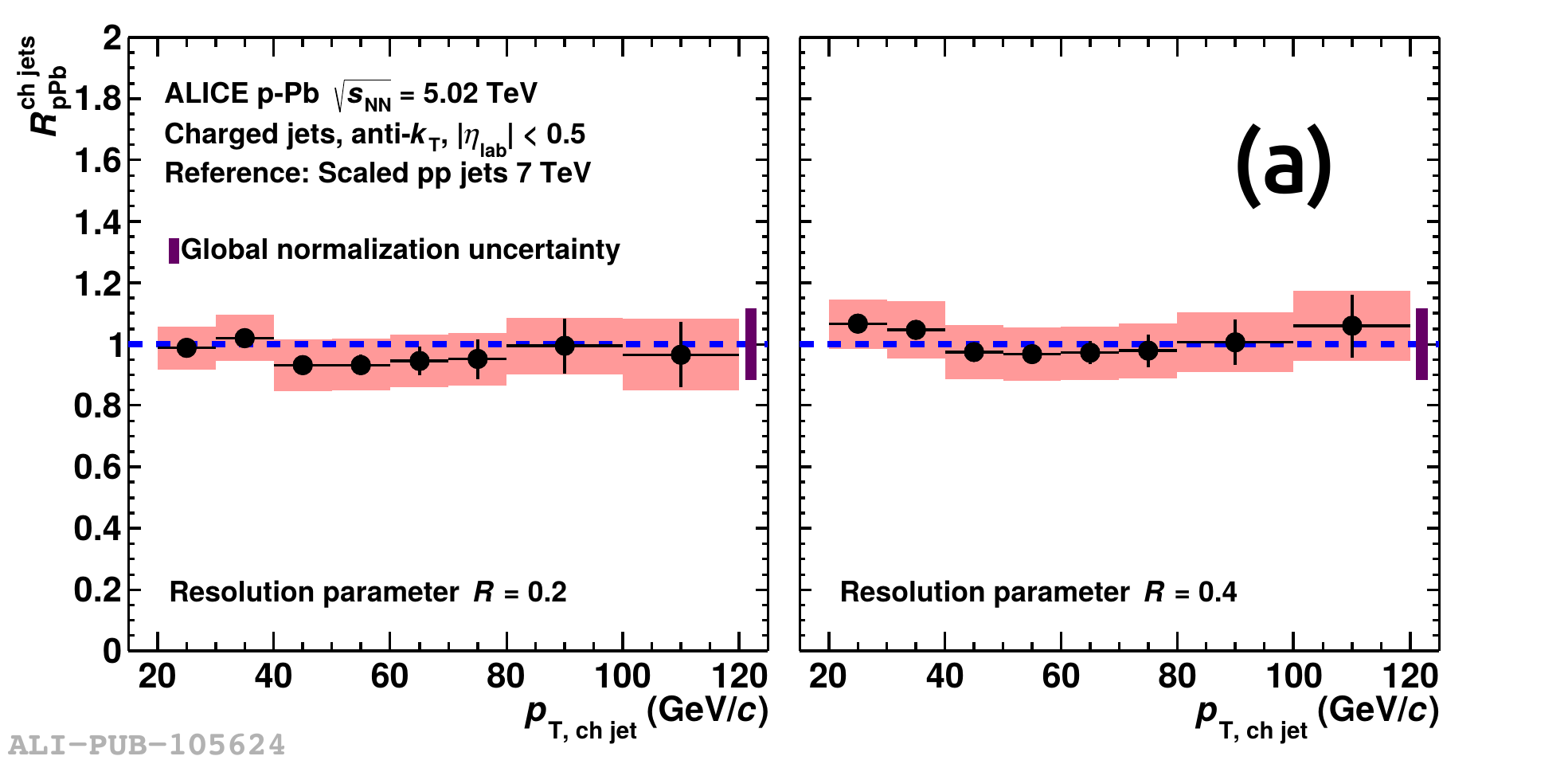}
\includegraphics[width=0.49\textwidth]{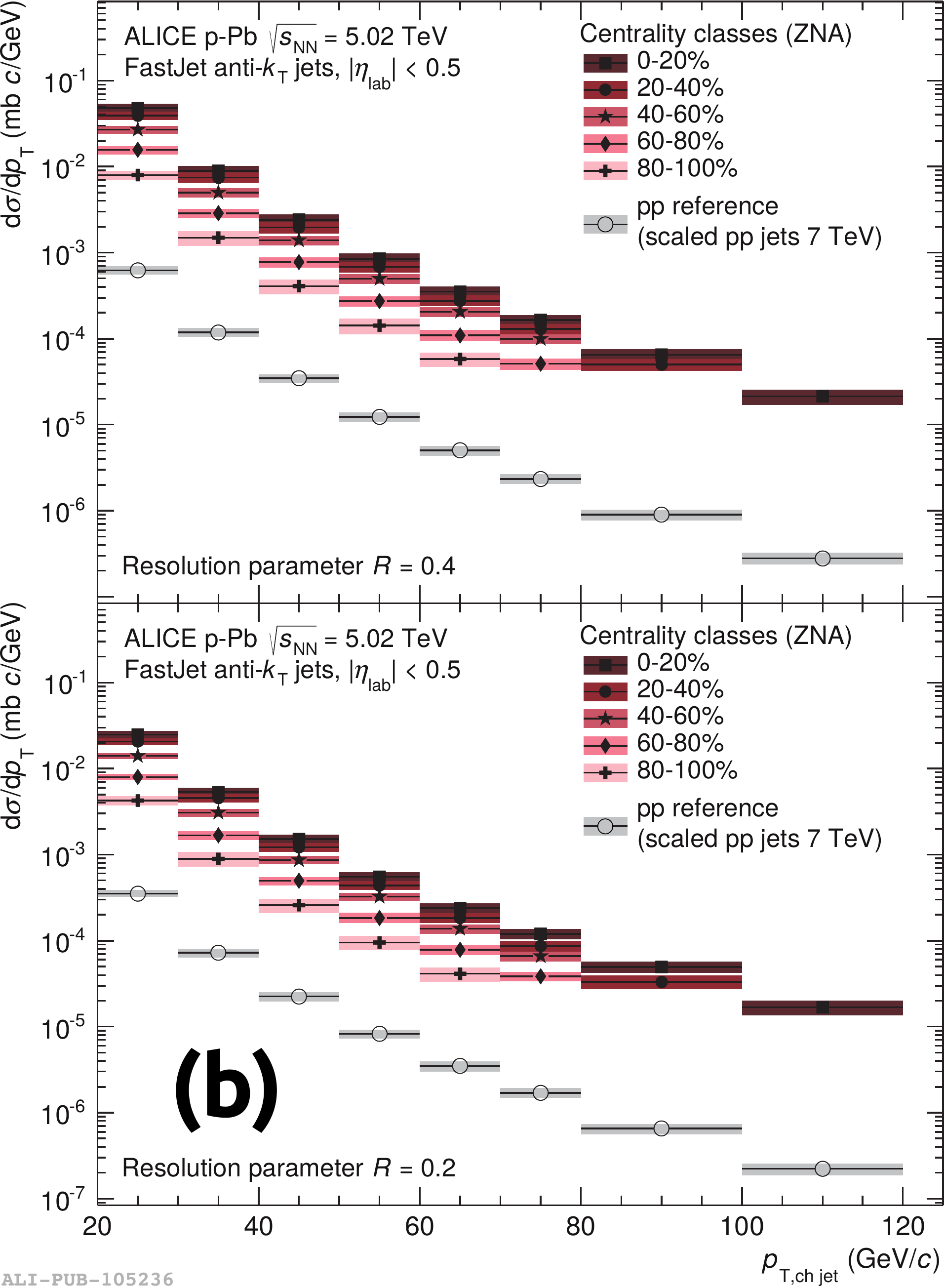}
\includegraphics[width=0.49\textwidth]{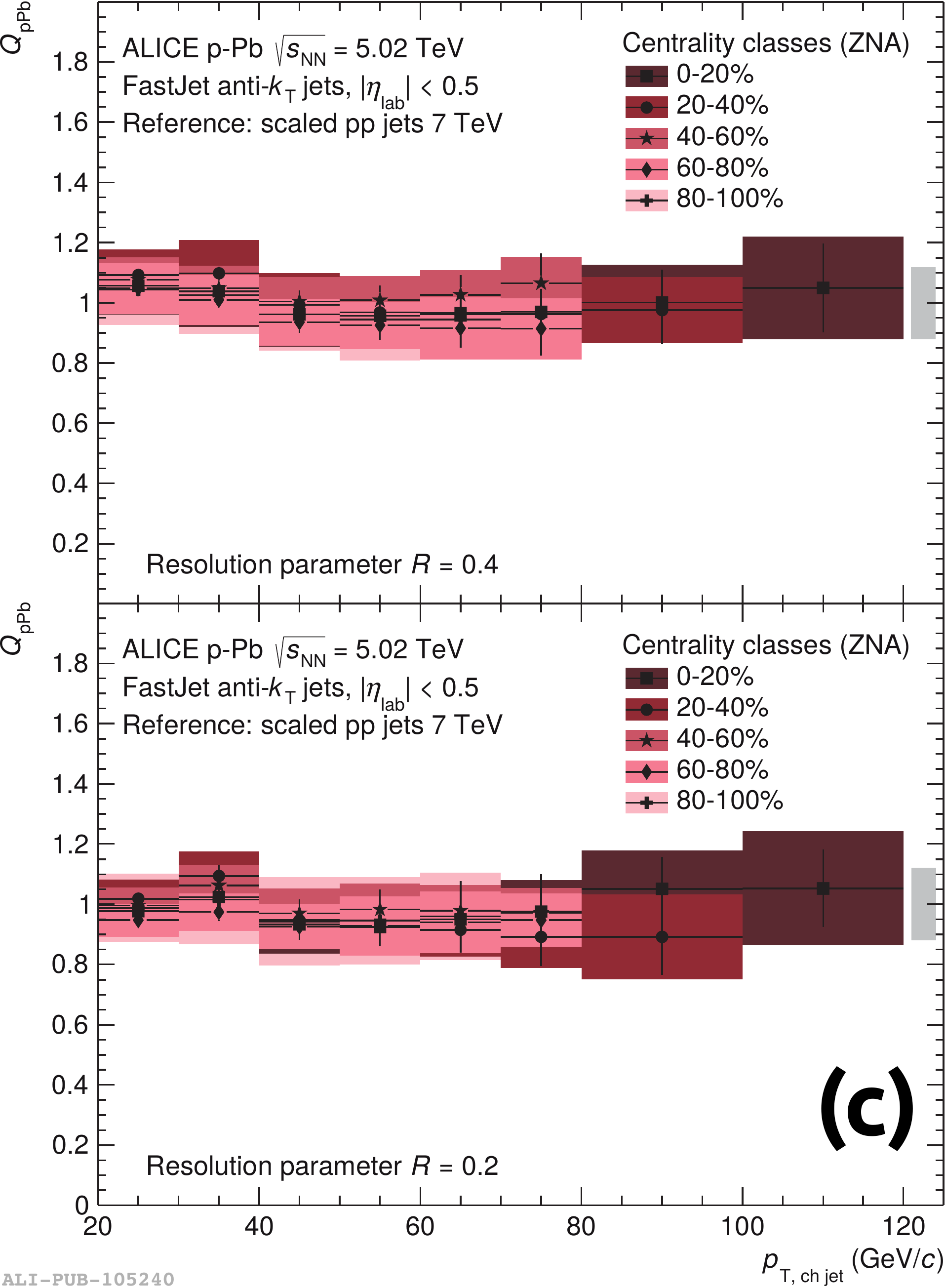}
\centering
\caption{(a) Charged jet nuclear modification factor for $R=0.2$ and $R=0.4$ for minimum bias p--Pb collisions. (b) Centrality-dependent charged jet cross sections in p--Pb collisions. Note that the spectra are not corrected for the number of binary collisions. (c) Centrality-dependent charged jet nuclear modification factor for $R=0.2$ and $R=0.4$ for p--Pb collisions.}
\label{fig:Panel1}
\end{figure}

\begin{figure}
\includegraphics[width=0.49\textwidth]{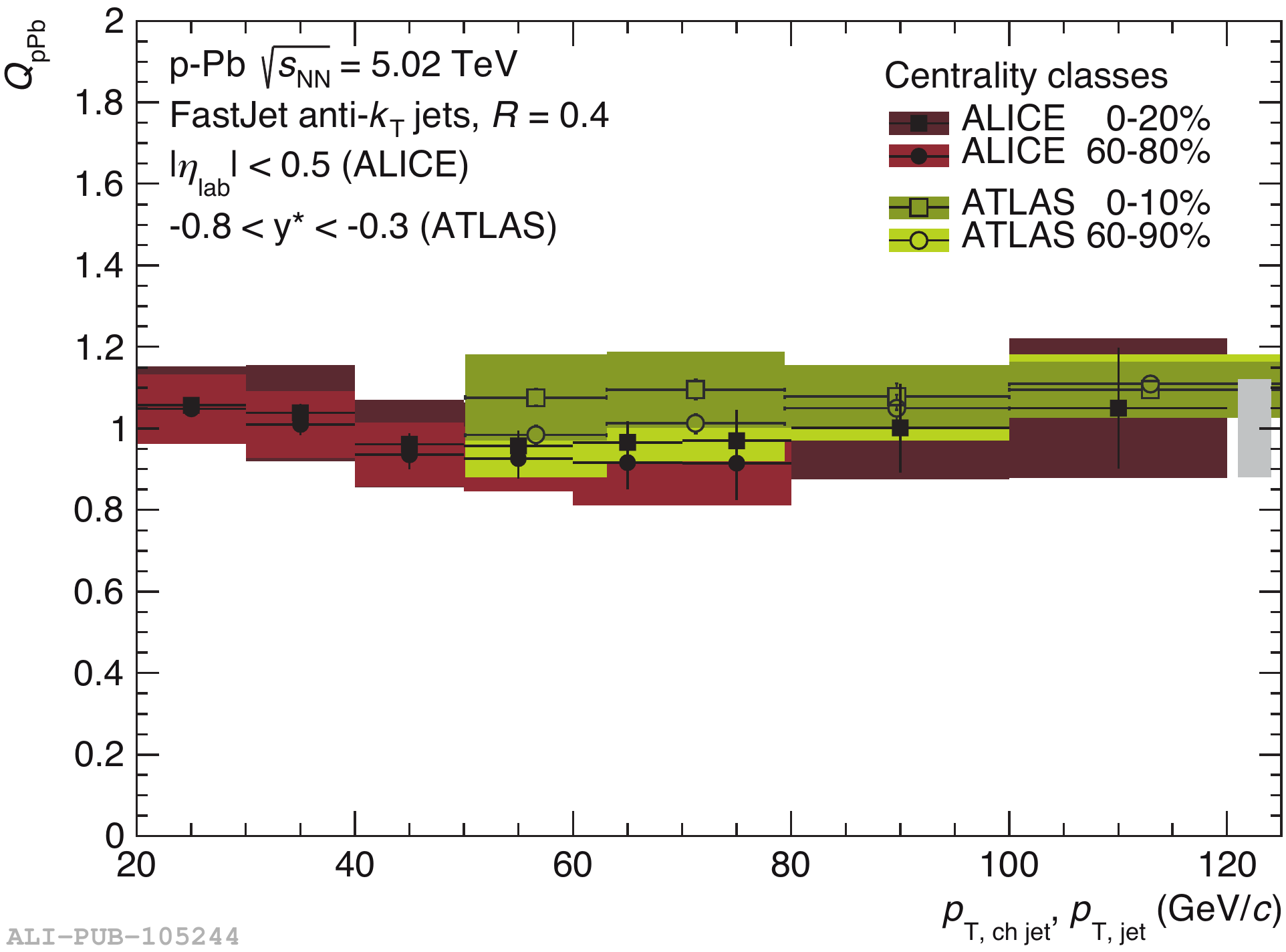}
\includegraphics[width=0.49\textwidth]{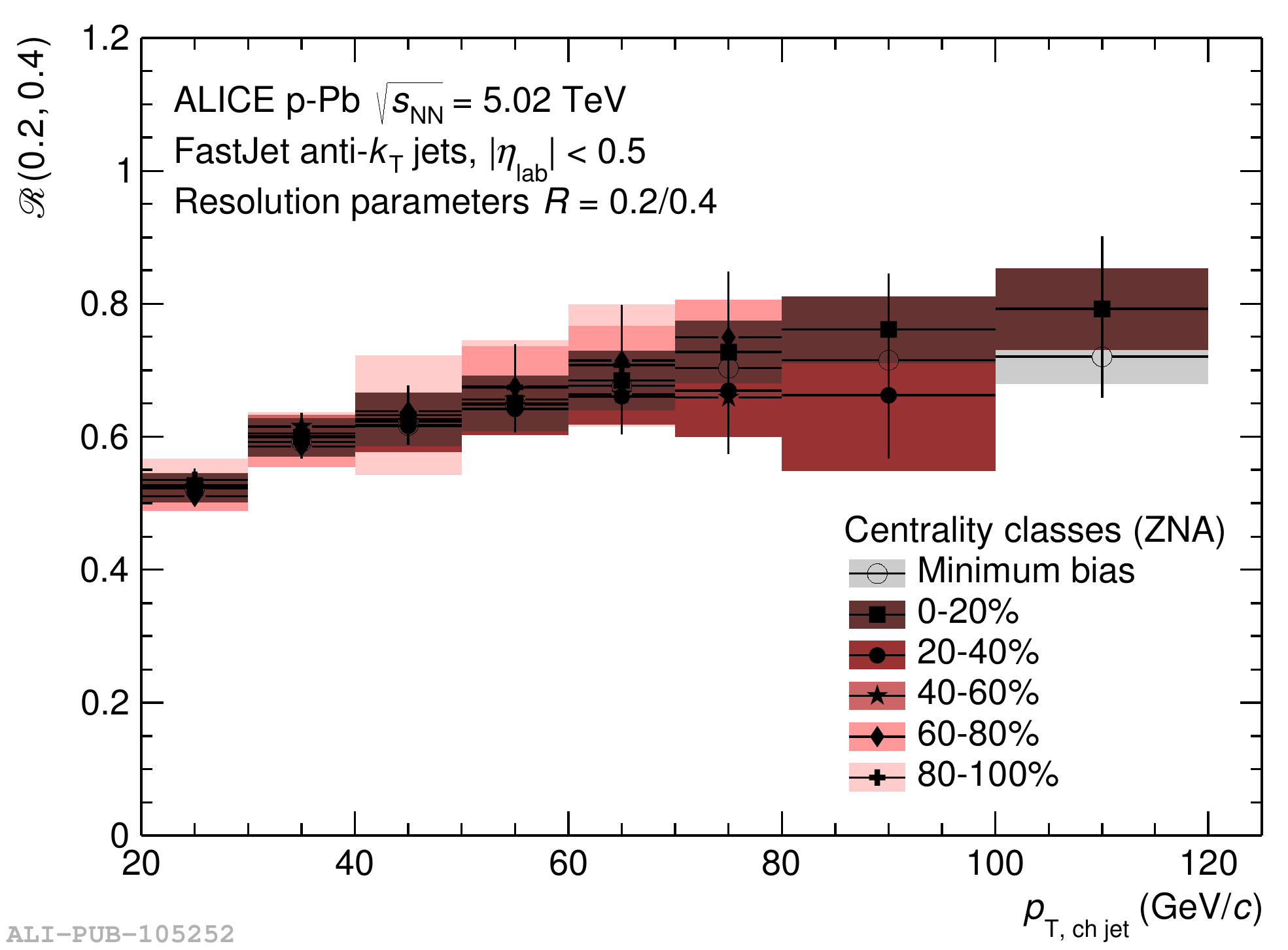}
\centering
\caption{Left: Comparison of the ALICE and ATLAS charged jet nuclear modification factors for $R=0.4$ and different centralities for p--Pb collisions. Right: Centrality-dependent jet cross section ratios $R=0.2/R=0.4$.}
\label{fig:Panel2}
\end{figure}

Figure \ref{fig:Panel1}a shows the charged jet nuclear modification factor for $R=0.2$ and $R=0.4$, respectively. It has been measured with 20 and 120 $\GeVc$ jet momentum and shows no significant modification within the uncertainties.  A good agreement with NLO pQCD calculations (not shown) is observed for both resolution parameters~\cite{MinBiasPaper}.\\

The centrality-dependent results are shown in Figs. \ref{fig:Panel1}b and c -- Fig. \ref{fig:Panel1}b shows the cross section measurement in bins of centrality, in Fig. \ref{fig:Panel1}c the nuclear modification factor is depicted in bins of centrality. Like for the minimum bias results, no significant modification and therefore also no centrality dependence has been observed in the measured range between 20 and 120 $\GeVc$. In addition, a comparison to ATLAS data on full jet measurements in Fig. \ref{fig:Panel2} (left) shows a good agreement in the comparable region of transverse momentum and pseudo rapidity.\\

As a very simple measure for the jet collimation, the jet cross section ratio was measured for $R=0.2/R=0.4$, see Fig. \ref{fig:Panel2} (right). Within the uncertainties, it shows no significant modification in bins of centrality. This observable was also calculated in pp collisions and several Monte Carlo simulations were performed (not shown here). A good agreement has been observed.

%___________________________________________________________________________________________
\section{Summary}
\label{sec:Summary}

In the two data analyses presented here, charged jets were measured in p-Pb collisions at $\snn = 5.02$ TeV at the LHC with the ALICE experiment both for minimum bias events and in bins of centrality. A good agreement of the minimum bias cross sections with NLO pQCD calculations was observed for both analyzed resolution parameters $R=0.2$ and $R=0.4$.

The nuclear modification factors show no significant effect, neither pointing to a strong nuclear modification nor to a strong centrality dependence. In addition, the result is compatible with the ATLAS full jet measurement.

The charged jet cross section ratio does not point to any (strong) change in the jet structure compared to pp collisions and several Monte Carlo simulations. Additionally, the ratio exhibits no significant centrality dependence.

%___________________________________________________________________________________________

%___________________________________________________________________________________________


\section*{References}

\begin{thebibliography}{99}

\bibitem{MinBiasPaper}
  ALICE Collaboration:
%  \emph{Measurement of charged jet production cross sections and nuclear modification in p-Pb collisions at $\sqrt{s_\mathrm{NN}} = 5.02$ TeV},
  \emph{Phys. Lett. B} {\bf 749} (2015) 68-81.

\bibitem{CentralityPaper}
  ALICE Collaboration:
%  \emph{Centrality dependence of charged jet production in p-Pb collisions at $\sqrt{s_\mathrm{NN}} = 5.02$ TeV},
  {\tt [nucl-ex/1603.03402]},
  \emph{accepted by EPJC}.

\bibitem{ALICE2008}
  ALICE Collaboration:
%  \emph{The ALICE experiment at the CERN LHC},
  \emph{JINST} {\bf{3}} (2008) 8002.

\bibitem{ALICE2010b}
  ALICE Collaboration:
%  \emph{The ALICE TPC, a large 3-dimensional tracking device with fast readout for ultra-high multiplicity events},
  \emph{Nucl. Instrum. Meth.} {\bf{A622}} (2010) 316-367.
%  {\tt [physics.Ins-det/1001.1950]}.

\bibitem{ALICE2010}
  ALICE Collaboration:
%  \emph{Alignment of the ALICE Inner Tracking System with cosmic-ray tracks},
  \emph{JINST} {\bf{5}} (2010) 3003.
%  {\tt [physics.Ins-det/1001.0502]}.

\bibitem{ALICE2010c}
  ALICE Collaboration:
%  \emph{Charged-Particle Multiplicity Density at Midrapidity in Central Pb--Pb Collisions at $\sqrt{s_\mathrm{NN}}=2.76$ TeV},
  \emph{Phys. Rev. Lett.} {\bf{105}} (2010) 252301.
%  {\tt [nucl-ex/1011.3916]}.

\bibitem{CentralityMethodPaper}
  ALICE Collaboration:
%  \emph{Centrality dependence of particle production in p-Pb collisions at $\sqrt{s_\mathrm{NN}}$ = 5.02 TeV},
  \emph{Phys. Rev. C} {\bf 91} (2015) 064905.
%  {\tt [nucl-ex/1412.6828]}.

\bibitem{Cacciari2008}
  M. Cacciari, G.P. Salam, and G. Soyez:
%  \emph{The anti-$k_T$ jet clustering algorithm},
 	\emph{JHEP} {\bf 0804} (2008) 063.
%  {\tt [hep-ph/0802.1189]}.

\bibitem{Cacciari2006}
  M. Cacciari and G.P. Salam:
%  \emph{Dispelling the $N^3$ myth for the $k_t$ jet-finder},
 	\emph{Phys. Lett. B} {\bf 641} (2006) 57-61.
%  {\tt [hep-ph/0512210]}.

%\bibitem{KTBackgroundCMS}
%  CMS Collaboration:
%  \emph{Measurement of the underlying event activity in pp collisions at $\sqrt{s} = 0.9$ and 7 TeV with the novel jet-area/median approach},
%  \emph{JHEP} {\bf 08} (2012) 130.
%  {\tt [hep-ex/1207.2392]}.

\bibitem{SVDAlgorithm}
  A. Hoecker and V. Kartvelishvili:
%  \emph{SVD Approach to Data Unfolding},
  \emph{Nucl. Instrum. Meth.} {\bf A372} (1996) 469-481.
%  {\tt [hep-ph/9509307]}.

%\bibitem{CentralityPbPb}
%  ALICE Collaboration:
%  \emph{Centrality determination of Pb--Pb collisions at $\sqrt{s_\mathrm{NN}}$ = 2.76 TeV with ALICE},
%  \emph{Phys. Rev. C} {\bf 88} (2013) 44909.
%  {\tt [nucl-ex/1301.4361]}.

%\bibitem{CrossSectionPaper}
%  ALICE Collaboration:
%  \emph{Measurement of visible cross sections in proton-lead collisions at $\sqrt{s_\mathrm{NN}} = 5.02$ TeV in van der Meer scans with the ALICE detector},
%  \emph{JINST} {\bf 9} (2014) 1100.
%  {\tt [nucl-ex/1405.1849]}.

%\bibitem{ALICE2014a}
%  ALICE Collaboration:
%  \emph{Measurement of charged jet suppression in Pb--Pb collisions at $\sqrt{s_\mathrm{NN}}=$ 2.76 TeV},
%  \emph{JHEP} {\bf 30} (2014) 013.
%  {\tt [nucl-ex/1311.0633]}.

%\bibitem{PYTHIA6}
%  T. Sjoestrand, S. Mrenna, and P. Skands:
%  \emph{PYTHIA 6.4 Physics and Manual},
%  \emph{JHEP} {\bf 05} (2006) 026.
%  {\tt [hep-ph/0603175]}.

%\bibitem{GEANT3}
%  R. Brun, F. Carminati, and S. Giani:
%  \emph{GEANT Detector Description and Simulation Tool},
%  \emph{CERN-W-5013} (1994).


\end{thebibliography}
\end{document}